\documentclass[twocolumn,aps,showpacs,prl,amsmath,amssymb,floatfix,superscriptaddress]{revtex4-2}

\usepackage{graphicx}
\usepackage{dcolumn}
\usepackage{bm}
\usepackage{physics} 
\usepackage{xcolor} 
\usepackage{bbm} 
\usepackage[colorlinks, hypertexnames=false]{hyperref} 

\renewcommand{\epsilon}{\varepsilon}
\renewcommand{\theta}{\vartheta}

\renewcommand{\vec}{\bm}

\begin{document}

\preprint{APS/123-QED}

\title{Exceptional Points and Optimized Encoding 
in Qubit Stabilizer Codes}

\author{Yves Rottstaedt}
\email{yves.rottstaedt@uni-leipzig.de}
\affiliation{%
Institut für Theoretische Physik, Universität Leipzig, 04103 Leipzig, Germany
}

\author{Andrei I. Pavlov}%
\affiliation{%
TKM, Karlsruhe Institute of Technology, 76131 Karlsruhe, Germany
}%
\author{Alexander Shnirman}%
\affiliation{%
TKM, Karlsruhe Institute of Technology, 76131 Karlsruhe, Germany
}%
\affiliation{%
IQMT, Karlsruhe Institute of Technology, 76131 Karlsruhe, Germany
}%
\author{Yuval Gefen}%
\affiliation{%
Department of Condensed Matter Physics, Weizmann Institute of Science, Rehovot 7610001, Israel
}%
\author{Bernd Rosenow}%
\affiliation{%
Institut für Theoretische Physik, Universität Leipzig, 04103 Leipzig, Germany
}%

\date{\today}

\begin{abstract}
Continuous blind stabilizer measurements turn stabilizer codes into a Liouvillian spectral problem in which Hamiltonian terms compete with dissipative error channels. We show that this competition generically enables Liouvillian exceptional points, whose occurrence at weak error rates is constrained by the code distance. Exceptional points are detrimental in this regime of error correction as they maximize the decay rate of encoded coherences. We then optimize the encoding of the logical qubit using the eigenmodes of the Lindblad operator. This phase-optimized encoding cancels the linear short-time decay of logical qubit coherences. We apply this encoding to the [[5,1,3]] code and demonstrate that it outperforms naive encoding methods near the exceptional point, improves the fidelity of a logical $\pi$ rotation and remains effective in the presence of additional weak error channels.
\end{abstract}

\maketitle

\emph{Introduction.---}
Protecting quantum  information from decoherence and control errors is a prerequisite for scalable quantum computation~\cite{terhal_quantum_2015,campbell_roads_2017}. Error-mitigation protocols can reduce the noise of near-term devices without implementing a full redundant encoding~\cite{temme_error_2017,li_efficient_2017}, but long-time storage and fault-tolerant processing require quantum error correction (QEC), in which a logical degree of freedom is embedded in a larger Hilbert space~\cite{nielsen_quantum_2010,terhal_quantum_2015}. Stabilizer codes provide the canonical framework for active QEC. They are often written in the form $[[N,k,d]]$, where $N$ is the number of physical qubits, $k$ the number of encoded qubits, and $d$ the code distance~\cite{terhal_quantum_2015}. Errors are inferred from stabilizer syndromes and are then corrected~\cite{shor_scheme_1995,steane_error_1996,gottesman_stabilizer_1997}, or equivalently tracked in a Pauli frame~\cite{knill_quantum_2005}. This active approach differs from passive protection, where information is placed in symmetry-protected or dynamically decoupled sectors, such as noiseless and decoherence-free subspaces or subspaces protected by pulse sequences~\cite{zanardi_noiseless_1997,lidar_decoherence-free_1998,viola_dynamical_1999,uhrig_keeping_2007}. The standard circuit description of QEC proceeds in discrete rounds of syndrome extraction and recovery. When syndrome information is obtained continuously, or when many weak operations are coarse grained, the dynamics is instead naturally described by a Lindblad master equation~\cite{gorini_completely_1976,lindblad_generators_1976,azouit_adiabatic_2016}. Continuous QEC has been developed theoretically and experimentally in several settings, including continuous error tracking~\cite{paz_continuous_1998,ahn_continuous_2002,van_handel_feedback_2005,lidar_continuous-time_2013,livingston_experimental_2022,mohseninia_always-on_2020,atalaya_continuous_2021}. In this description, coherent imperfections appear as spurious Hamiltonian terms, while stochastic errors and eliminated syndrome-changing coherent terms appear as jump processes. Such terms model realistic sources of error, including residual couplings, crosstalk, weak-parity-measurement imperfections, and physically motivated Lindblad noise~\cite{barenco_elementary_1995,denhez_quantum-error-correction_2012,krinner_realizing_2022,xu_zz_2021,wang_control_2022,novais_hamiltonian_2008,schwartzman-nowik_modeling_2025,halaski_obstacles_2026}.

The competition between coherent evolution and dissipation can produce exceptional points (EPs), non-Hermitian degeneracies at which both eigenvalues and eigenmodes coalesce~\cite{heiss_physics_2012,miri_exceptional_2019}. In open quantum systems, EPs occur in the Liouvillian spectrum and are shaped by quantum jumps and the Hamiltonian time evolution~\cite{minganti_quantum_2019,khandelwal_emergent_2025,wu_experimental_2026}. They can be useful for steering and state preparation, because the Liouvillian gap is maximized at an EP and therefore ensures fast relaxation into the steady sector~\cite{kumar_optimized_2022,zhou_accelerating_2023}. For QEC, however, the same gap enhancement is potentially harmful. It accelerates relaxation of encoded coherences and thereby makes the choice of encoding dynamically important. 

In this Letter, we report that: 1) the possibility of Liouvillian EPs is a generic feature of stabilizer codes under blind syndrome measurement (the detector readouts are traced out). 2) The occurrence of EPs at weak error rates is constrained by the code distance. Normally, the EPs can emerge in stabilizer codes only at substantially high error rates. However, due to undetectable spurious Hamiltonian terms, the EPs may occur at arbitrarily weak error rates. 3) To remedy this vulnerability of the stabilizer codes to EPs, we identify a syndrome-conditioned encoding of the logical qubit that cancels the linear short-time decay of coherences and thereby suppresses the EP-related enhancement
of the decoherence. We first demonstrate the mechanism for the $[[5,1,3]]$ Laflamme--Miquel--Paz--Zurek (LMPZ) code~\cite{laflamme_perfect_1996,bennett_mixed-state_1996}, then generalize it to arbitrary stabilizer codes, and finally construct the optimized encoding.

\emph{Exceptional points in stabilizer codes.---}
We consider a stabilizer code consisting of N physical qubits. For the sake of clarity, we consider codes that encode one logical qubit, though the presented arguments of this section can be straightforwardly generalized for an arbitrary $[[N, k, d]]$ code with $k\geq 1$. For a code with a single logical qubit, we have $N-1$ commuting stabilizer generators $S_k=S_k^\dagger$, $S_k^2=I$, and logical Pauli operators $X_L,Z_L$. The stabilizer eigenvalues label syndrome sectors. Continuous blind measurement of the stabilizers suppresses coherences between different syndrome sectors~\cite{ahn_continuous_2002}. In the strong-measurement, coarse-grained limit, Hamiltonian terms that commute with all stabilizers act within syndrome sectors, while non-commuting terms create rapidly dephased inter-sector coherences that can be eliminated from the Hamiltonian and are described as effective Lindbladian dissipator terms~\cite{azouit_adiabatic_2016}. We therefore only consider Hamiltonians that commute with all stabilizer generators and jump operators that anticommute with at least one generator. Assuming one dominant detected error channel $E$, we describe the time evolution using the master equation
\begin{equation}\label{Liouvllian}
        \dot{\rho} = \mathcal{L}\rho = -i[H,\rho] + \Gamma \sum_{k=1}^{N - 1} \mathcal{D}[S_k]\rho + \gamma \mathcal{D}[E]\rho \, ,
\end{equation}
with
\begin{equation}
    \mathcal{D}[J]\rho = J\rho J^\dagger - \frac{1}{2} \left\{ J^\dagger J,\rho \right\} \, .
\end{equation}
Here $\Gamma$ is the measurement-induced dephasing rate and $\gamma$ is the rate of the error channel. To suppress the coherence between the syndrome sectors, we take the strong measurement limit where $\Gamma$ is much larger than all other energy scales.

The Liouvillian operator $\mathcal{L}$ (\ref{Liouvllian}) may possess EPs in the long-living modes of its spectrum. Let us clarify this mechanism in {detail} using a concrete example. For that, we use the $[[5,1,3]]$ LMPZ code, whose stabilizer generators are
\begin{align}
\label{eq:stabilizers}
 S_1=XZZXI, \qquad S_2=IXZZX,  \nonumber \\
 S_3=XIXZZ, \qquad S_4=ZXIXZ,
\end{align}
and whose logical operators can be chosen as
\begin{equation}
 X_L=XXXXX,  \; Z_L=ZZZZZ, \; Y_L=YYYYY.
\end{equation}
Let us assume there is an error channel described by $E=IXIII$, which only anticommutes with $S_1$ therefore the system switches between the sectors with $s_1=+1$ and $s_1=-1$. Let us choose a Hamiltonian which contains a logical Pauli operation and an additional spurious term, 
\begin{equation}\label{eq:hamiltonian}
    H = \frac{\Omega}{2} X_L + \frac{\epsilon}{2} X_L S_1.
\end{equation}
This Hamiltonian implements the logical $X_L$ rotation, while the system has an additional small amplitude term $X_LS_1$, $\epsilon\ll \Omega$. The Hamiltonian spurious term has Pauli weight three and commutes with all stabilizers. Physically, the spurious Hamiltonian terms can appear when implementing the logical Hamiltonian gates. In this case, the logical multi-qubit operation is decomposed into multiple gates on fewer qubits. Effects such as crosstalk can then lead to such Hamiltonian spurious terms~\cite{berke_transmon_2022}.

The spurious term in Eq. (\ref{eq:hamiltonian}) anticommutes with the error operator, $\{E, X_LS_1\}=0$ and commutes with the logical drive, $[X_L, X_LS_1]=0$. This is sufficient to produce EPs in the Liouvillian spectrum of this system (\ref{Liouvllian}). As $E$ only anticommutes with $S_1$, only this stabilizer is affected by the errors. Fixing $s_2=s_3=s_4=1$ leaves us with a description of two syndrome sectors (corresponding to $\pm$ signs of the first stabilizer syndrome). As we operate in the strong measurement limit  $\Gamma \gg \gamma,\Omega,\epsilon$, the stabilizer measurements suppress coherences between sectors, but not within them and we are left with an $8\times8$ Liouvillian block describing the dynamics of the two syndrome sectors exactly.

\begin{figure}[t]
\includegraphics[width=\linewidth]{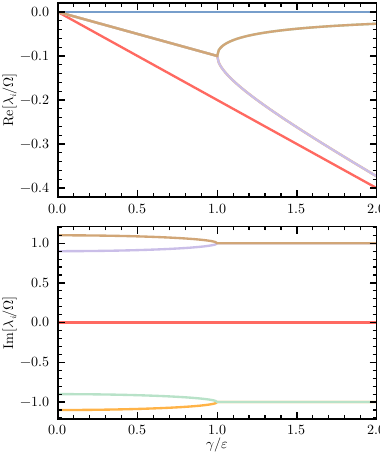}
\caption{\label{fig:spectrum} Spectrum of the Liouvillian for the [[5,1,3]] code with the Hamiltonian Eq.~\eqref{eq:hamiltonian} and the error $E=IXIII$ in the subspace of two syndrome sectors, using $\epsilon = 0.1$ and $\Omega = 1.0$. The upper panel shows the real part and the lower panel shows the imaginary part. Stationary and fast decaying modes are each double degenerate and have no imaginary part. The remaining four modes form two conjugate pairs and give two second-order EPs at $\gamma = \epsilon$.}
\end{figure}

Figure~\ref{fig:spectrum} shows the spectrum of this block. There are two steady modes and two fast decaying modes, each without imaginary part, and two complex-conjugate pairs whose eigenvalues coalesce at $\gamma=\epsilon$. We therefore get two EPs with complex conjugate eigenvalues. This stems from the fact that the Liouvillian preserves hermiticity of the density matrix and the EPs come from coherence modes which are the hermitian conjugate of each other. Thus, the EPs will always have the same critical error rate and cannot be split up. The $8\times8$ Liouvillian decomposes into $L_\mathrm{sub}=\mathrm{diag}(L_\mathrm{ss},L_\mathrm{EP})$ with $L_\mathrm{ss}$ containing the steady and fast decaying modes, respectively. 
The $4\times4$ block $L_{EP}$ containing an EP acts in the subspace of Liouvillian modes defined by the basis of four operators $B=\{ P_+ \Sigma_+,P_-\Sigma_+,P_+ \Sigma_-,P_-\Sigma_- \}$, where $P_\pm$ projects onto the sector with $s_1=\pm1$ and $\Sigma_\pm=(Z_L\pm iY_L)/2$. In this basis, the block containing EPs can be written as
\begin{equation}
\label{eq:Lred} \resizebox{.9\hsize}{!}{$
 L_\mathrm{EP} =
 \begin{pmatrix}
   i (\Omega + \epsilon) & -\gamma & 0 & 0 \\
 -\gamma &  i (\Omega - \epsilon) & 0 & 0 \\
 0 & 0 &   - i (\Omega + \epsilon) & -\gamma \\
 0 & 0 & -\gamma &  - i (\Omega - \epsilon) \\
 \end{pmatrix}
 - \gamma \mathbbm{1}_4 .$}
 \end{equation}
The $L_{ss}$ block has two steady eigenmodes $e_1$ and $e_2$ and two other fast decaying modes, which are no longer used. The second steady mode is the conserved operator $X_L$, which commutes with $H$ and $E$; such a conserved, possibly stabilizer-dressed logical operator exists whenever the weak-rate EP conditions below hold. The EP block has eigenmodes denoted $e_3,e_4,e_5,e_6$. At $\gamma=\epsilon$, the pairs $e_3,e_4$ and $e_5,e_6$ coalesce, producing generalized eigenmodes $\tilde e_3$ and $\tilde e_5$. The explicit construction of these modes is provided in the {Supplemental} Material \cite{Suppl}.

The physical nature of these EPs can be understood by noting that the spurious term $X_LS_1=s_1X_L$ makes the drive amplitude syndrome dependent, $\Omega\to\Omega+s_1\epsilon$, while the error $E$ flips $s_1$ at rate $\gamma$. A similar phenomenon occurs in the problem of dephasing of a qubit by a random-telegraph fluctuator~\cite{paladino_decoherence_2002,galperin_non-gaussian_2006,bergli_decoherence_2009,paladino_1f_2014}.

We now identify \textit{sufficient} conditions for EPs at arbitrarily weak error rates in an arbitrary stabilizer code.
Every admissible Hamiltonian term is a logical operator dressed by stabilizers,
\begin{align}\label{eq:all_hams}
    O\left( \{ \mu_k \} \right) = O_L \prod_{k=1}^{N-1}(S_k)^{\mu_k}, \, \mu_k = 0,1,
\end{align}
with $O_L\in \{X_L,Y_L,Z_L\}$. Formally, $O_L$ could also be the identity, which however only shifts the whole spectrum within each stabilizer sector. Any other term does not commute with at least one stabilizer and therefore is suppressed by the high rate stabilizer measurements. Let us consider an arbitrary Hamiltonian consisting of three distinct types of Pauli terms: $H=H_L+H_C+H_A$, with $H_L$ responsible for logical operation(s), $H_C$ spurious terms that commute with the error $E$ ($[H_C,E]=0$), and $H_A$ spurious terms that anticommute with $E$ ($\{H_A,E\}=0$). The sufficient conditions for the emergence of an EP at  small $\gamma$ are  $[H_L,E]=0$ and $[H_A,H_C+H_L]=0$. The basic principle is that EPs necessarily emerge through the anticommutation of the Hamiltonian terms with the dissipative error operators. As long as a detectable error $E$ anticommutes with the Hamiltonian, it mixes two syndrome sectors whose energy shifts have exactly the same amplitude but opposite signs. The resulting pair of modes forms an effective non-Hermitian two mode problem, cf. Eq.~\eqref{eq:LEP2}, with an EP emerging when the energy shift and the dissipation rate match, $\epsilon=\gamma$. 
\begin{equation}\label{eq:LEP2}
    \tilde{L} = \begin{pmatrix}
        -\gamma-i\epsilon & \gamma \\
        \gamma & -\gamma + i\epsilon
    \end{pmatrix},
\end{equation}
The EP emerges here even if these two parameters are arbitrarily small. The example of Eq.~\eqref{eq:Lred} is a special case of this general principle; see End Matter for details.

Let us estimate how often these requirements are satisfied in various situations. This is a lower-bound estimate based on sufficient conditions for EPs to emerge. We can consider three distinct scenarios of increasing complexity: (i) the Hamiltonian consists of a single term of Eq.~\eqref{eq:all_hams}. This term either commutes ($H_A=0$, no EP) or anticommutes ($H_C=0$, EP) with the error. Since at least one stabilizer $S_i$ anticommutes with the error $E$, and $S_i^2=I$, 
for each member
of the set~\eqref{eq:all_hams} that commutes with E there is a partner that anticommutes.
 Thus, exactly half of all terms in \eqref{eq:all_hams} anticommute with $E$ and produce an EP of the type~\eqref{eq:LEP2} at arbitrarily weak error rates.
(ii) the Hamiltonian consists of several such terms but no logical operation term,  $H=H_C+H_A$. In this case, the sufficient condition for an EP is the commutation between this Hamiltonian contribution, $[H_C, H_A]=0$ \footnote{strictly speaking, there are fine-tuned exceptions from this rule, as is explained in the End Matter, but they have measure zero weight in a continuum of the system parameters since they require exact cancellation between certain amplitudes in the Hamiltonian}. (iii) the Hamiltonian consists of a logical Pauli term and spurious terms ($H=H_L+H_C+H_A$): The energy scale of $H_L$ far exceeds realistic error inverse time scales, so weak-rate EPs require the logical term to commute with the error, $[H_L,E]=0$. To generate an EP the anticommuting spurious part must commute with the rest of the Hamiltonian: $[H_A,H_C+H_L]=0$ (the system in Eq. \eqref{eq:hamiltonian} is an example of this case). The fraction of spurious terms generating an EP is code-specific. Let us assume that the Hamiltonian consists of exactly two terms. For the LMPZ code, the fraction of such terms (out of all admissible terms) when averaging over all possible detectable errors is $39/94$ for the case (ii) and $11/126$ for the case (iii) (see End Matter).

The fact that all Hamiltonian terms defined by Eq.~\eqref{eq:all_hams} commute with all stabilizers, constrains the terms that can produce weak-rate EPs through the code distance. By the definition of the code distance $d$, every undetectable Pauli operator that acts nontrivially on the logical qubit has weight at least $d$, so the same holds for all spurious terms of the form of Eq.~\eqref{eq:all_hams}. Assuming that terms with smaller Pauli weight are more likely, this implies that codes with larger distance are better protected against EPs. Conversely, if the measurement of a stabilizer generator fails, previously detectable low-weight terms become undetectable, exposing a bigger risk for EPs; a discussion of both aspects for the LMPZ code is given in the End Matter. A stabilizer code without undetectable spurious Hamiltonian terms is protected against EPs up to error rates set by the scale of the logical operations.

\emph{Optimized Encoding.---}
Instead of limiting our consideration to a single-syndrome sector (all $s_i=+1$), we can construct a qubit using Liouvillian modes of Eq.~\eqref{eq:Lred} from two sectors. In the two-syndrome-sector subspace, any encoded qubit state can be represented as
\begin{equation}\label{eq:state_rep}
    \rho = \frac{1}{4} \left( \eta_0 + \vec{r} \cdot \vec{\eta} \right),
\end{equation}
where $|\vec r|\leq1$, $\eta_0$ is the identity on the reduced subspace, and $\eta_i$ obey the Pauli algebra on that subspace. An encoding is therefore a choice of the operator basis $\eta_i$. Because the dynamics is generated by $\mathcal L$, it is natural to construct this basis from Liouvillian eigenmodes which offers the advantages described below.

We set $\eta_0=e_1$ and $\eta_3=e_2$, so that the logical $z$ axis is associated with the two steady modes. The remaining Pauli operators are chosen from the EP block. At the EP we start with the regular eigenmodes and add the generalized eigenmodes with free amplitude and phase while assuring hermiticity as
\begin{align}\label{eq:encoding_EP}
        \eta_1(\theta,r) &= e_3 + e_5  + r e^{i \vartheta} \tilde{e}_3 + r e^{-i \vartheta} \tilde{e}_5 \, , \nonumber \\
        \eta_2(\theta,r) &= -ie_3 +ie_5 -i r e^{i \vartheta} \tilde{e}_3 + i r e^{-i \vartheta} \tilde{e}_5  \, . 
    \end{align}
Enforcing the Pauli algebra fixes the amplitude to $r=2\epsilon\sin\theta$ and leaves a single phase $\theta$ free such that we can define $\eta_i^{(\theta)} = \eta_i(\theta,r(\theta))$. In the encoded plane, $e^{\mathcal{L}t}$ acts as the unitary evolution generated by the effective logical Hamiltonian $H_\mathrm{eff}=\Omega\eta_3/2$ times a scalar decay function, which at the EP reads $f_\theta(t)=e^{-\epsilon t}\left[1+\sin(2\theta)\,\epsilon t\right]$. The component orthogonal to the plane, generated by the spurious term, is annihilated by the logical readout and does not affect the fidelities calculated below at $t_0=\pi/\Omega$. The remaining freedom in $\theta$ therefore controls only the loss of encoded coherence: $\theta=\pi$ retains only the two slowest decaying modes and gives the naive encoding, $f_\pi(t)=e^{-\epsilon t}$, whereas $\theta=\pi/4$ cancels the linear short-time decay and defines the optimized encoding. Away from the EP, the same construction is obtained by replacing each generalized eigenmode by the corresponding eigenmode difference, since $e_3-e_4\propto\tilde e_3$ and $e_5-e_6\propto\tilde e_5$, with the naive phase continued analytically below the EP and the optimized phase again fixed by canceling the linear term (End Matter). For the present model, the optimized Pauli operators are independent of $\gamma$ and take the simple form $\eta_1^\mathrm{opt}=-Z_LS_1$ and $\eta_2^\mathrm{opt}=Y_LS_1$ which commute with the error operator. These are the logical operators conjugated by $E$ in the flipped sector, i.e., the syndrome-conditioned Pauli frame of error tracking~\cite{knill_quantum_2005,knill_theory_2000,kribs_unified_2005}, which the slowest-mode prescription~\cite{macieszczak_towards_2016} misses near the EP because the right eigenmodes are strongly non-orthogonal there. Dressing logical operators with stabilizer generators is also used in work on error-transparent subspaces~\cite{vy_error-transparent_2013,young_error_2013,ma_error-transparent_2020,kapit_error-transparent_2018}.

We quantify the effect of this encoding by a logical $\pi$ rotation. The initial state is the $+1$ eigenstate of $\eta_1$, $\rho_i=(\eta_0+\eta_1)/4$, the target is the $-1$ eigenstate $\rho_t=(\eta_0-\eta_1)/4$ reached after $t_0=\pi/\Omega$, and the performance is measured by the mixed-state fidelity of $\rho_t$ and $\rho_f=e^{\mathcal{L}t_0}\rho_i$~\cite{jozsa_fidelity_1994}, $F=[\Tr\sqrt{\sqrt{\rho_t}\rho_f\sqrt{\rho_t}}]^2$. To compare encodings we use the infidelity-reduction factor $G=(1-F_\mathrm{naive})/(1-F_\mathrm{opt})$.
Because of the two-sector encoding, the initial and final state are already mixed. In practice, one would need to prepare a pure state where the unwanted coherences decay exponentially with $\Gamma$ due to the stabilizer measurements. The effect on the fidelity is negligible and discussed in Ref.~\cite{Suppl}.

\begin{figure}[t]
    \centering
    \includegraphics[width=\linewidth]{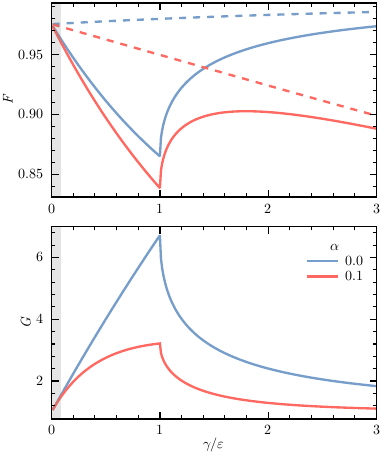}
    \caption{$\pi$ rotation starting in $\rho_i =(\eta_0 + \eta_1)/4$ using $\epsilon = 0.1$ and $\Omega = 1.0$ for both one (blue) and two error channels (red). The top panel shows the fidelity for the optimized (dashed lines) and naive (solid lines) encodings, and the bottom panel shows the infidelity-reduction factor. In the shaded region, an encoding using only one syndrome (calibrated to its sector frequency, see text) yields higher fidelity.}
    \label{fig:fid_all}
\end{figure}

Figure~\ref{fig:fid_all} shows in blue the fidelity of the $\pi$ rotation for the naive ($F_\mathrm{naive}$, solid) and optimized ($F_\mathrm{opt}$, dashed) encoding and displays below the infidelity-reduction factor. The optimized encoding outperforms the naive encoding over the whole range of possible error rates, but is especially pronounced at the EP. Moreover, the naive encoding behaves similarly to the Liouvillian gap in the two-sector subspace. Indeed, we can express the fidelity for the naive encoding as
\begin{equation}\label{eq:Fgap}
        F_\mathrm{naive} = \frac{1 + \cos \left(\mathrm{Re}\left[ \sqrt{\epsilon^2 - \gamma^2} \right] t_0\right) e^{-\Delta t_0}}{2} \, ,
\end{equation}
with the Liouvillian gap $\Delta = - \max_{\lambda_i \neq 0} \mathrm{Re}\,\lambda_i$ in the two-sector subspace.
At the EP, $1 - F_\mathrm{naive} = (1 - e^{-\epsilon t_0})/2 \approx \epsilon t_0/2$ while $1 - F_\mathrm{opt} = (1 - (1+\epsilon t_0)e^{-\epsilon t_0})/2 \approx (\epsilon t_0)^2/4$, so $G \approx 2/(\epsilon t_0)$: the advantage grows as the spurious coupling weakens. For general $\gamma$, $f_\mathrm{opt}(t)$ (End Matter) shares the asymptotic decay rate of $f_\mathrm{naive}$, so the gain is confined to times $t\lesssim1/\epsilon$, and $F_\mathrm{opt}(\gamma)$ is smooth across the EP.
One effect of encoding the qubit within two syndrome sectors is that the fidelity does not approach one as we send the error strength to zero. This stems from the fact that for $\gamma<\epsilon$ we have two coherent frequencies, $\Omega\pm\epsilon$: the pulse $t_0=\pi/\Omega$ over- and under-rotates the two sectors by $\pm\pi\epsilon/\Omega$, a coherent error that sets the floor $F=\cos^2(\pi\epsilon/2\Omega)\approx0.975$. Thus, an encoding using only one syndrome sector could perform better. In this case, one fixes all syndrome bits and is left with a two-dimensional subspace for the encoding, with $t_0=\pi/(\Omega+\epsilon)$ calibrated to the sector frequency. The fidelity then approaches one in the limit of vanishing error rate. The shaded region in Fig.~\ref{fig:fid_all} shows where this single-sector encoding performs better than the optimized encoding, the border being at $\gamma_\mathrm{crit}/\epsilon \approx 0.09$. 

We assumed before that one detectable error channel is dominant. However,  it is natural to expect that other channels might be present. To test the robustness of the encoding scheme against additional noise, we add an error channel $E_\mathrm{add}=IZIII$ with rate $\gamma_\mathrm{add}=\alpha\gamma$. This channel couples the optimized two-sector subspace to additional syndrome sectors. We again compute the fidelity for a $\pi$ rotation according to the full dynamics for $\alpha=0.1$. The increase of the fidelity at large $\gamma$ (Fig.~\ref{fig:fid_all}, upper panel, blue dashed line) is an artifact of the idealized scenario when strictly one error is present in the system. Even a small additional error results in the suppression of the fidelity at large values of $\gamma$, as one would expect. While fidelity is overall reduced, we still see that the phase-optimized encoding outperforms the naive encoding. The EP survives in the spectrum in this case, even though this additional error does not commute with the logical Hamiltonian, stressing that the spectral EPs in stabilizer codes are actually even more common than one could expect from the lower-bound estimates.

\emph{Conclusion.---}
We have shown that Liouvillian EPs in stabilizer codes {arise} generically because detectable errors commonly anticommute with stabilizer-compatible Hamiltonian terms, whose Pauli weight is bounded from below by the code distance. At such points, the same gap enhancement that can accelerate dissipative state preparation maximizes the decay rate of eigenmode-based encoded coherences. Encoding the logical qubit across two syndrome sectors in the syndrome-conditioned Pauli frame cancels the linear short-time decay and mitigates the loss of fidelity at and around the EP, also under weak additional error channels. Both the EPs and their remedy are features of continuous QEC with blindly measured stabilizer generators, and the encoding carries over to other stabilizer codes.

\emph{Acknowledgements.--}
We acknowledge funding from the German Research Foundation, Grant No. 536777080.

\bibliography{refs}

\appendix 
\onecolumngrid \vspace{0.5cm}
\begin{center}
{\large \bf End Matter}\\
\end{center}
\twocolumngrid

\emph{EPs follow from anticommutation.---}
We first show that anticommutation between a Hamiltonian and a detected error is sufficient for a Liouvillian EP to be possible and then comment on the case of only partially anticommuting Hamiltonians. Syndrome sectors are labeled by
\begin{equation}
    \bm{s} = (s_1,...,s_{N-1}), \quad s_k=\pm1,
\end{equation}
and the projectors onto the sectors are
\begin{equation}\label{eq:projector}
    P_{\bm{s}} = \prod_{k=1}^{N-1} \frac{I+s_kS_k}{2}.
\end{equation}
A detectable Pauli error maps one syndrome sector to another, $EP_{\bm{s}}E=P_{\bm{s}'}$. If, in addition, the error anticommutes with the Hamiltonian, $EHE=-H$, then the same error connects states with opposite Hamiltonian energies. Equivalently, for eigenstates satisfying $H\ket{\bm s,\pm}=\pm\epsilon\ket{\bm s,\pm}/2$, one can choose phases such that $E\ket{\bm s,\pm}=\ket{\bm s',\mp}$.

The relevant Liouvillian modes are the off-diagonal density-matrix elements within each syndrome sector. Their evolution is
\begin{align}
    \mathcal{L} \ketbra{\bm s, \pm}{\bm s, \mp} &= \mp i\epsilon \ketbra{\bm s,\pm}{\bm s, \mp} \nonumber \\
    &+ \gamma(\ketbra{\bm s^\prime, \mp}{\bm s^\prime, \pm} - \ketbra{\bm s,\pm}{\bm s, \mp}) .
\end{align}
The first term is the coherent phase accumulation, while the second term describes stochastic switching between the two syndrome sectors. In the basis $\{\ketbra{\bm s,+}{\bm s,-},\ketbra{\bm s',-}{\bm s',+}\}$ this gives
\begin{equation}\label{eq:LEP}
    L = \begin{pmatrix}
        -\gamma-i\epsilon & \gamma \\
        \gamma & -\gamma + i\epsilon
    \end{pmatrix}.
\end{equation}
The two eigenvalues are $-\gamma\pm\sqrt{\gamma^2-\epsilon^2}$ and coalesce with their eigenvectors at $\gamma=|\epsilon|$. Thus anticommutation is a sufficient condition for a second-order Liouvillian EP.

Considering Hamiltonians with anticommuting and commuting parts like Eq.~\eqref{eq:hamiltonian} can be analyzed equivalently. We call terms that commute with error operator even and those that anticommute odd. In order to have a closed subspace in which we find the second order EP as in Eq.~\eqref{eq:LEP}, the even and odd parts have to commute. Then, we get Liouvillian blocks of the form
\begin{align}
    L_\mathrm{gen} = \begin{pmatrix}
        -\gamma-i\epsilon_1 & \gamma \\
        \gamma & -\gamma + i\epsilon_2
    \end{pmatrix},
\end{align}
where $\epsilon_1$ and $\epsilon_2$ can differ by sign and magnitude. Those Liouvillians also have an EP at $\gamma = |\epsilon_1 + \epsilon_2|/2$. There are special cases where $\epsilon_1 = - \epsilon_2$ that do not have EPs, but these are fine-tuned rather than generic.

\begin{figure}[t]
    \includegraphics[width=\linewidth]{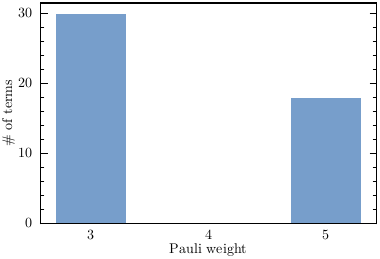}
    \caption{\label{fig:ham_hist} Number of possible Hamiltonian terms sorted by their Pauli weight for the LMPZ code.}
\end{figure} 
\begin{figure}[t] 
    \includegraphics[width= \linewidth]{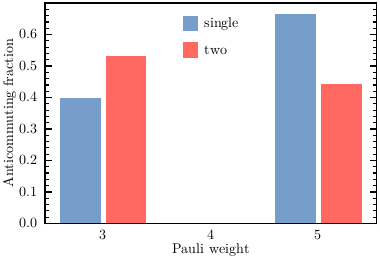}
    \caption{\label{fig:anti_hist}Number of anticommuting pairs of Hamiltonian and error terms sorted by the Pauli weight of the Hamiltonian term for the LMPZ code. The histogram differentiates between single- (blue) and two-qubit (red) errors. Shown is the fraction of (term, error) pairs that anticommute, for terms of the given weight: $2/5$ and $2/3$ (single-qubit), $8/15$ and $4/9$ (two-qubit) for weights three and five.}
\end{figure} 

\emph{Counting argument and application to LMPZ code.---}
To estimate how frequently the EP conditions from the main part are fulfilled, we count the pairs of Hamiltonian terms and error operators that satisfy the relevant commutation relations. Starting with case (i) of the main text, which corresponds to the Hamiltonian comprised of a single term, we require this Hamiltonian term to anticommute with the error operator. All non-trivial Hamiltonian terms can be written in the form of Eq.~\eqref{eq:all_hams}. For the LMPZ code, the total number of terms is $2^4 \cdot 3 = 48$. We can further resolve those by their Pauli weight, which is shown in Fig.~\ref{fig:ham_hist}. There we can see that all terms have a Pauli weight of at least three, as the code has distance three. Moreover, the majority of terms have a Pauli weight of three, while the remaining terms have a Pauli weight of five. The detectable errors comprise all single- and two-qubit Pauli errors. The counting argument works for every fixed detectable error which we can choose to be $E=IXIII$ for the current example. This error operator only anticommutes with $S_1$. Thus, multiplication by $S_1$ maps Hamiltonian terms that commute with the error operator to those that anticommute. The simplest example would be the two terms from the Hamiltonian in Eq.~\eqref{eq:hamiltonian}. $X_L$ commutes with the error operator while $S_1X_L$ anticommutes with it. Since the stabilizer generators are unitary and self-adjoint, $S_k^2=I$, multiplication by $S_k$ defines a one-to-one pairing of commuting and anticommuting terms. Therefore, for each fixed detectable error, exactly one half of all stabilizer-compatible Pauli Hamiltonian terms anticommute with it. The anticommuting fraction obtained by averaging over all detectable errors is consequently $1/2$. Namely, $24$ of the $48$ possible Hamiltonian terms anticommute with the error operator and produce an EP in the spectrum. If we assume that the measurement of the stabilizer generator $S_1$ is faulty, this leads to more possible Hamiltonian terms. The terms are the errors which only change the first syndrome bit and thus commute with every other stabilizer. They include one single-qubit error, namely $IXIII$ the one studied in the main part, and six two-qubit errors. 

Although the total anticommuting fraction derived above is $1/2$, this fraction can be additionally resolved with respect to the Pauli weight of both Hamiltonian and error terms. Figure~\ref{fig:anti_hist} shows the anticommuting fraction as a function of Hamiltonian weight and separates single- from two-qubit errors. While a higher Pauli weight leads to a higher anticommuting fraction for single-qubit errors, this is reversed for two-qubit errors.

Moving on to case (ii) of the main text, we assume that the Hamiltonian consists of two spurious terms. We choose these two terms randomly and independently (they may not coincide). We again start by fixing an error operator. Then, we draw two Hamiltonian terms and first have to check if they anticommute with the error operator. When both anticommute with the error, we have an EP. If only one of them anticommutes with the error, then they have to commute with each other to guarantee an EP. Lastly, if both commute with the error there is no EP in the spectrum. Averaging over all possible Hamiltonians and error operators for the LMPZ code shows that the fraction of combinations which yield an EP is $39/94$.

Finally, we can perform the counting for case (iii) of the main text. Here, we assume for simplicity that the Hamiltonian consists of a logical operator and a spurious term. We again fix the error operator and can look at the example of $E=IXIII$. In order to get weak-rate EPs, $E$ has to commute with the logical operator. The only logical operator commuting with the error is $X_L$. Now, we can count the spurious terms that commute with $X_L$ and anticommute with $E$. In the present example $8$ out of $45$ possible spurious terms (the dressed terms with $\mu\neq0$) fulfill those conditions. However, to get the total fraction of possible Hamiltonians giving an EP, we have to average over all logical operators and errors. As not every logical operator is compatible with every error operator, the total fraction of possible combinations yielding an EP for the LMPZ code is $11/126$.

\emph{Dynamics of the encoded operators.---}
At the EP, enforcing the Pauli algebra fixes the amplitude to $r=2\epsilon\sin(\theta)$. The time-dependent operators in the encoded plane are
\begin{align}\label{eq:eta_t}
        \eta_1^{(\theta)} (t) = P_{||}[e^{\mathcal{L}t}\eta_1^{(\theta)}] = \left[ \cos(\Omega t) \eta_1^{(\theta)} + \sin(\Omega t) \eta_2^{(\theta)}\right] f_\theta (t), \nonumber \\
        \eta_2^{(\theta)} (t) = P_{||}[e^{\mathcal{L}t}\eta_2^{(\theta)}] = \left[ \cos(\Omega t) \eta_2^{(\theta)} - \sin(\Omega t) \eta_1^{(\theta)}\right] f_\theta (t),
\end{align}
where $P_{||}$ is the projector onto the encoded plane. At the EP, we get
\begin{equation}\label{eq:f_theta}
        f_\theta(t)=e^{-\epsilon t}\left[1+\sin(2\theta)\epsilon t\right].
\end{equation}
Thus the nontrivial Liouvillian dynamics decomposes into the desired logical rotation generated by $H_\mathrm{eff}=\Omega\eta_3/2$ and a scalar decay. Choosing $\theta=\pi/4$ removes the linear term in the short-time expansion of $f_\theta(t)$, defining the optimized encoding. The naive encoding corresponds to the phase $\theta_\mathrm{naive}=\pi$, i.e., to using only the two slowest decaying modes, and decays exponentially. Away from the EP, with $\kappa = \sqrt{\gamma^2 - \epsilon^2}$, the corresponding decay functions are
\begin{align}\label{eq:f_decay}
        f_\mathrm{naive}(t) &=  \mathrm{Re}\left[e^{-\left(\gamma-\kappa\right)t}\right], \nonumber \\
        f_\mathrm{opt}(t) &=  e^{-\gamma t}  \left[ \cosh \left( \kappa t \right) + \frac{\gamma}{\kappa} \sinh \left( \kappa t \right) \right] \nonumber \\
        &= 1 - \frac{\epsilon^2 t^2 }{2} + \mathcal{O}\left( t^3 \right).
\end{align}
The absence of a linear term is the defining feature of the optimized encoding and explains its enhanced fidelity for short and intermediate logical rotations.

\emph{Operational aspects.---}
Operationally, the encoding requires knowledge of the dominant error operator, but of neither $\gamma$ nor $\epsilon$. A pure initial state suffices: the state in sector $(+1,1,1,1)$ with $Z_L = -1$ is a $+1$ eigenstate of $\eta_1^\mathrm{opt}$, and components outside the encoded plane are invisible to the logical readout. Readout is the weight-three Pauli measurement $\eta_1^\mathrm{opt} = YIIYZ$, equivalently a joint measurement of $S_1$ and $Z_L$. Since the encoding is a syndrome-conditioned choice of Pauli frame, it composes with standard active decoding for all other syndromes.

\newpage
\onecolumngrid
\begin{center}
{\large \bf Supplementary Material}\\
\end{center}

\emph{Construction of optimized encoding.---}
We give the explicit construction of the reduced Liouvillian and the optimized encoding. For our particular Hamiltonian the choice of syndrome bits $s_2,s_3,s_4$ is not relevant because the Hamiltonian has the same action in each two syndrome sector subspace. For more complicated Hamiltonians this choice may affect the position of the exceptional points and also needs to be optimized. The two syndrome sectors used in the main text are $P_\pm=P_{(\pm1,1,1,1)}$, with projectors
\begin{equation}\label{eq:projector}
    P_{\bm{s}} = \prod_{k=1}^{N-1} \frac{I+s_kS_k}{2}.
\end{equation}
A convenient operator basis for the full reduced Liouvillian is
    \begin{align}
        B = \{  &P_+, P_-, P_+ X_L, P_- X_L, P_+ \Sigma_+, P_- \Sigma_+, P_+ \Sigma_-, P_- \Sigma_- \}.
    \end{align}
In this basis, the Liouvillian restricted to the two-sector subspace is
\begin{equation}
        L_{\mathrm{sub}}=
        \begin{pmatrix}
            -\gamma & \gamma & 0 & 0 & 0 & 0 & 0 & 0 \\
            \gamma & -\gamma & 0 & 0 & 0 & 0 & 0 & 0 \\
            0 & 0 & -\gamma & \gamma & 0 & 0 & 0 & 0 \\
            0 & 0 & \gamma & -\gamma & 0 & 0 & 0 & 0 \\
            0 & 0 & 0 & 0 &-\gamma + i (\Omega + \epsilon) & -\gamma & 0 & 0 \\
            0 & 0 & 0 & 0 & -\gamma & -\gamma + i (\Omega - \epsilon) & 0 & 0 \\
            0 & 0 & 0 & 0 &0 & 0 &   -\gamma - i (\Omega + \epsilon) & -\gamma \\
            0 & 0 & 0 & 0 &0 & 0 & -\gamma &  -\gamma - i (\Omega - \epsilon) \\
        \end{pmatrix}.
    \end{equation}
The first four basis elements form the steady-state and fast-decay block, while the last four basis elements form the EP block.

The eigenvalues and corresponding eigenmodes are
    \begin{align} 
        \lambda_{1/2} &= 0 \, , \quad \lambda_{3/4} = -i \Omega - \gamma \mp \sqrt{\gamma^2 - \epsilon^2}  , \nonumber \\
        \lambda_{5/6} &= i \Omega - \gamma \mp \sqrt{\gamma^2 - \epsilon^2} \, , \quad \lambda_{7/8} = -2\gamma \, ,
    \end{align}
with
    \begin{align}
        e_1 &= (P_+ + P_-),  \nonumber \\
        e_2 &= X_L (P_+ + P_-), \nonumber \\
        e_{3/4} &= \Sigma_- \left( \frac{i \epsilon \pm \sqrt{\gamma^2 -\epsilon^2}}{\gamma} P_+ + P_- \right), \nonumber \\
        e_{5/6} &=  \Sigma_+ \left( \frac{-i \epsilon \pm \sqrt{\gamma^2 -\epsilon^2}}{\gamma}  P_+ + P_- \right), \nonumber \\
        e_7 &= (P_+ - P_-), \nonumber \\
        e_8 &= X_L (P_+ - P_-).
    \end{align}
At the EP, the coalesced eigenmodes are supplemented by the generalized eigenmodes
    \begin{align}
    \tilde{e}_3 = -\Sigma_- P_+/\epsilon, \nonumber \\
    \tilde{e}_5 = -\Sigma_+ P_+/\epsilon.
    \end{align}
They satisfy the Jordan-chain relations $L_{\mathrm{sub}}\tilde e_{3/5}= \lambda_{3/5} \tilde e_{3/5} + e_{3/5}$ and are the natural directions along which the encoding can be varied at fixed Pauli algebra.

We define $\eta_0=e_1$ and $\eta_3=e_2$. The remaining logical Pauli operators are parameterized at the EP by
    \begin{align}\label{eq:encoding_EP_SM}
        \eta_1(\theta,r) &= e_3 + e_5  + r e^{i \vartheta} \tilde{e}_3 + r e^{-i \vartheta} \tilde{e}_5 \, , \nonumber \\
        \eta_2(\theta,r) &= -ie_3 +ie_5 -i r e^{i \vartheta} \tilde{e}_3 + i r e^{-i \vartheta} \tilde{e}_5  \, .  
    \end{align}
Requiring $\eta_i\eta_j=\delta_{ij}\eta_0+i\epsilon_{ijk}\eta_k$ fixes the amplitude to $r=2\epsilon\sin(\theta)$ and leaves the phase $\theta$ as the only free \ br{parameter,} allowing us to define $\eta_i^{(\theta)} = \eta_i(\theta,r(\theta))$.

The construction extends away from the EP by replacing generalized eigenmodes with eigenmode differences. This works as $e_3-e_4\propto\tilde e_3$ and $e_5-e_6\propto\tilde e_5$. Again enforcing the algebra yields
\begin{align}\label{eq:encoding}
    \eta_1(\theta,r) &=\left[ (e_3+e_4) +(e_5 + e_6) + r e^{i \vartheta} (e_3-e_4) + \mathrm{sgn}(\gamma - \epsilon) r e^{-i \vartheta} (e_5-e_6) \right]/2 \, , \nonumber \\
    \eta_2(\theta,r) &=\left[ -i(e_3+e_4) +i(e_5 + e_6) -i r e^{i \vartheta} (e_3-e_4) + \mathrm{sgn}(\gamma - \epsilon) i r e^{-i \vartheta} (e_5-e_6) \right]/2 \, ,  \nonumber \\
    r &= \frac{-\epsilon \cos(\vartheta - \frac{\Theta (\gamma- \epsilon)\pi}{2}) + \sqrt{\gamma^2 - \epsilon^2 \sin^2 \left(\vartheta - \frac{\Theta (\gamma- \epsilon)\pi}{2}\right)}}{\sqrt{|\gamma^2-\epsilon^2|}} \, .
\end{align}
The naive encoding corresponds to the phase $\theta_\mathrm{naive}=\pi$, i.e., to using only the two slowest decaying modes. The optimized phase is
    \begin{equation}
        \theta_{\mathrm{opt}} = \pi +\mathrm{sgn}(\gamma - \epsilon) \arctan\left[ \left(\epsilon / \gamma\right)^{\mathrm{sgn}(\gamma - \epsilon)} \right] \, .
    \end{equation}
The time evolution of the encoded operators and the resulting decay functions $f_\theta(t)$, $f_\mathrm{naive}(t)$, and $f_\mathrm{opt}(t)$ are given in the End Matter of the main text. \\

\emph{Finite measurement strength.---}
So far, we assumed to be in the strong measurement limit such that all coherences between subspaces decay reasonably fast. When taking a finite value for the measurement rate, the initial state for the $\pi$ rotation will be a pure state with decaying coherences between the syndrome sectors. To estimate the additional error, we have to take a look at the spectrum of the full Liouvillian.

\begin{figure}[t]
    \centering
    \includegraphics[width=0.5\linewidth]{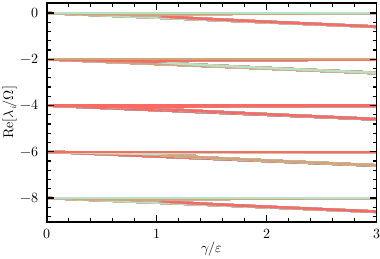}
    \caption{\label{fig:full_spectrum} Real part of the spectrum of the full Liouvillian for the [[5,1,3]] code, using $\epsilon = 0.1$, $\Omega = 1.0$, $\Gamma = 1.0$. All eigenvalues corresponding to coherences between syndrome sectors have an eigenvalue of at most $-2\Gamma$.}
    \label{fig:fid_all}
\end{figure}

The real part of the spectrum is shown in Fig.~\ref{fig:full_spectrum}. One can clearly identify five similar looking groups of eigenvalues. The uppermost group shows all eigenvalues inside the syndrome sectors, while all other groups are affected by at least one stabilizer measurement suppressing the dynamics. The slowest decaying coherences between syndrome sectors thus decay exponentially with a rate of $2\Gamma$.

Using this result, we can estimate a reduction of the fidelity when starting with a pure state instead of the idealized initial state used in the main text. Assuming all coherences between syndrome sectors to decay with rate $2\Gamma$, this yields a correction in fidelity of $\exp[-4 \Gamma t_0]$, where $t_0 = \pi/\Omega$ is the computation time for the $\pi$ pulse. Assuming $\Gamma = \Omega = 1.0$ results in a correction of $\exp[-4 \pi] \approx 3.5 \times 10^{-6}$, which is much smaller than the effect of intra-syndrome sector decoherences.

\end{document}